\documentclass{article}
\usepackage{graphicx}
\usepackage{amsmath}

\input{tcilatex}
\begin{document}

\title{Estimation for Unit Root Testing\thanks{%
Preliminary version not to be quoted without permission. Comments are
welcome.}}
\author{Dimitrios V. Vougas\thanks{%
School of Management, Accounting and Finance, Haldane Building, Room 26,
Singleton Park, Swansea SA2 8PP, UK. Tel. 0044-(0)1792-602102 (direct line).
Fax. 0044--(0)1792-295872 (departmental fax). E-mail: \textsf{%
D.V.Vougas@swan.ac.uk},\textsf{\ }Home Page: \textsf{%
http://www.swan.ac.uk/economics/staff/dv.htm}} \\
%EndAName
Swansea University}
\maketitle

\begin{abstract}
\noindent We revisit estimation and computation of the Dickey Fuller (DF)
and DF-type tests. Firstly, we show that the usual one step approach, based
on the "DF autoregression", is likely to be subject to misspecification.
Secondly, we clarify a neglected two step approach for estimation of the DF
test. (In fact, we introduce a new two step DF autoregression.) This method
is always correctly specified and efficient under the circumstances.
However, it is either neglected or misused in unit root testing literature.
The commonly employed hybrid of the (correct) two step method is shown to be
inefficient, even asymptotically. Finally, we further improve/robustify the
proposed two step method by employing the missing initial observations. Our
finally proposed method is to be used in unit root testing, since it is a
new DF autoregression that retains the missing observations.\newline
\newline
\newline
\textbf{Keywords:} Linear regression; Autoregressive error; Deterministic
component; Dickey Fuller autoregression; Two step autoregression; Unit root.%
\newline
\newline
\newline
\textbf{JEL classification:} C12, C13, C15.
\end{abstract}

\section{Introduction}

\noindent Econometricians are often accused of trying to discover
electricity by playing the radio. In the case of estimation for unit root
testing, some econometricians seriously believe they have discovered
electricity. That is they believe that their estimation methods are genuine,
fully efficient, and that they do not rely on previous literature.
Unfortunately, these beliefs are not true in general. It turns out, that
existing estimation methods for unit root testing quite likely misuse
existing estimation methods, or, even worse, efficient methods of the
literature are neglected. In general less, than needed, attention is paid to
estimation. Jansson and Nielsen (2012) and DeJong et al. (1992 a) clearly
imply the main estimation problem for unit root testing. This is estimation
of a linear, in the parameters, regression with autocorrelated error. Of
course, this is a problem a suitable version of the Cochrane-Orcutt (CO)
method, that allows for AR($p$) (ignoring the first $p$ (say) observations),
can handle. Or in its place any similar, fully iterated Gauss-Newton (GN)
(or any other) algorithm. However, due to the nature of employed regressors
(purely deterministic), one round CO type methods are suitable for
estimation and unit root inference. Dickey and Fuller (DF) (1979) are
clearly aware of the estimation problem, and their autoregression solves
this problem computationally very cheaply.\footnote{%
See also Nelson and Plosser (1982), Dickey and Said (1981), Said and Dickey
(1984, 1985), and Fuller (1996).} This is what we call the one step approach.%
\footnote{%
In effect, this method owes to Durbin (1960).} However, the DF approach is
liable to misspecification, and can be very easily used for con business!
Especially, if the deterministic component is anything but a trend
polynomial. In addition, there is no obvious way to amend the DF approach to
incorporate the first $p$ missing values.\newline
\newline
In this paper, we clarify and recommend a neglected estimation method that
calculates the DF and DF-type tests correctly. It is immune to
misspecification and any potential "saucy" business. This is the two step
method, we discuss and fully develop. It relies on work by Durbin (1970),
and subsequent research by Breusch (1978), and Godfrey (1978 a, b) and
(1988). No one has used this method for unit root testing before.\footnote{%
Or when used, an inefficient version of the estimation method is employed.}
The two step method has certain advantages. It guarantees correct
specification of the deterministic component, and can provide inference
about the structural parameters of the data generating process (DGP). In
addition, by proving the proposed method, we expose an inefficient version
of the method which is used in the literature. This method crudely
calculates the "DF" test from the least squares (LS) residuals of the DGP.
Such a resulting "DF" test is an inefficient variant of the original DF
test. The DF test exhibits particularly low power, and size distortion that
can be large in certain cases, see Schwert (1989), Agiakloglou and Newbold
(1992), DeJong et al. (1992 b). More recent studies on the DF test include
amongst others Leybourne \textit{et al}. (1998), Leybourne and Newbold
(1999, 2000), and Harvey \textit{et al}. (2009). In view of these findings,
it is of importance to amend the proposed two step estimation method in line
with findings in Belsley (1996). Our proposed amended two step method, uses
zero padded lags, so that the first $p$ rows of the DGP are implicitly used.
This new approach increases estimation efficiency and power, and robustifies
the resulting efficient DF test with respect to a neglected break. In
addition, the amended procedure alleviates the size problems of the original
DF test. This new DF test is to be called the efficient DF test.\newline
\newline
The paper is organised as follows: Section 2 discusses potential pitfalls
arising from estimation using the one step DF\ autoregression, while Section
3 shows an alternative two step DF autoregression, which is corrected to
retain the first observations. Finally, Section 4 concludes.

\section{Pitfalls of the One Step DF Autoregression}

\noindent An observed time series $y_{t}$ is generated via a deterministic
component ($x_{t}$) and a stochastic process ($z_{t}$)%
\begin{equation}
y_{t}=\gamma ^{\prime }x_{t}+z_{t}\text{, \ \ \ \ }t=1,...,T\text{.}  \tag{1}
\end{equation}%
The model is linear in the parameters. In addition%
\begin{equation}
z_{t}=\alpha z_{t-1}+u_{t}\text{,}  \tag{2}
\end{equation}%
with%
\begin{equation}
u_{t}=\xi (L)\varepsilon _{t}\text{, }\xi (L)=\sum_{i=0}^{\infty }\xi
_{i}L^{i}\text{, }\xi _{0}=1\text{, }\sum_{i=0}^{\infty }i|\xi _{i}|<\infty 
\text{, }\xi (1)\neq 0\text{.}  \tag{3}
\end{equation}%
$z_{0}$ is either an unknown constant or stochastically bounded, $O_{p}(1)$,
and $\varepsilon _{t}$ is a martingale difference sequence with $%
E\varepsilon _{t}^{2}=\sigma ^{2}$ and $\sup_{t}E\varepsilon _{t}^{4}<\infty 
$, see Stock (1991). The inverse of $\xi (L)$, say $b(L)$, $b(L)=\xi
(L)^{-1} $, assuming it exists, is approximated by a truncated AR($k$)
polynomial%
\begin{eqnarray}
b_{k}(L)u_{t} &=&\varepsilon _{kt}\text{, }b_{k}(L)=1-b_{1}L-...-b_{k}L^{k}%
\text{,}  \TCItag{4} \\
u_{t} &=&\sum_{j=1}^{k}b_{j}u_{t-j}+\varepsilon _{kt}\text{, }\varepsilon
_{kt}=\varepsilon _{t}+\sum_{j=k+1}^{\infty }b_{j}u_{t-j}  \notag
\end{eqnarray}%
with some decay in $k$, say $T^{-1/3}k\rightarrow 0$ as both $T$ and $k$
increase.\ Chan and Park (2002) discuss alternative truncation orders. From
Eq. (2) and (4), the long AR($p$) ($p=k+1$) representation for $z_{t}$ is%
\begin{equation}
z_{t}=\sum_{j=1}^{p}\rho _{j}z_{t-j}+\varepsilon _{kt}\text{, }t=p+1,...,T%
\text{,}  \tag{5}
\end{equation}%
where%
\begin{equation}
\rho (L)=1-\rho _{1}L-...-\rho _{k}L^{k}-\rho _{k+1}L^{k+1}=(1-\alpha L)b(L)%
\text{.}  \tag{6}
\end{equation}%
Note that $\rho (1)=0$, if and only if $\alpha =1$.\footnote{%
Approximately, $\varepsilon _{t}$ and $\varepsilon _{kt}$ have similar
properties.}\newline
\newline
The commonly used one step approach to calculate the DF\footnote{%
See Fuller (1996) and Dickey and Fuller (1979, 1981), Nelson and Plosser
(1982), Dickey and Said (1981), and Said and Dickey (1984, 1985).} or
DF-type tests combines Eq. (1) and (5), and employs the so-called DF
transformation. It owes to Durbin (1960), and derived by lagging Eq. (1), $%
j=1,...,p$ times, multiplying each resulting equation by $\rho _{j}$, and
subtracting each outcome from Eq. (1). The analysis below in effect extends
the work of DeJong et al. (1992 a). That is, one employs%
\begin{equation}
y_{t}=\sum_{j=1}^{p}\rho _{j}y_{t-j}+\gamma ^{\prime }x_{t}-\rho _{1}\gamma
^{\prime }x_{t-1}-...-\rho _{p}\gamma ^{\prime }x_{t-p}+\varepsilon _{kt}%
\text{, }t=p+1,...,T\text{.}  \tag{7}
\end{equation}%
Eq. (7) is always correctly specified, and for feasibility the term $\gamma
^{\prime }x_{t}-\rho _{1}\gamma ^{\prime }x_{t-1}-...-\rho _{p}\gamma
^{\prime }x_{t-p}$ must be correctly expanded. This may not be an easy task
in general. For $x_{t}=\{1,t,...,t^{r}\}^{\prime }$ (full $r$-th order
polynomial trend, with no power missing), with $\gamma =\{\gamma _{0},\gamma
_{1},...,\gamma _{r}\}^{\prime }$ being the vector of associated parameters,
we obtain%
\begin{equation}
y_{t}=\sum_{j=1}^{p}\rho _{j}y_{t-j}+\mu ^{\prime }x_{t}+\varepsilon
_{t}=\rho y_{t-1}+\sum_{j=1}^{k}\beta _{j}\Delta y_{t-j}+\mu ^{\prime
}x_{t}+\varepsilon _{kt}\text{, }t=p+1,...,T\text{.}  \tag{8}
\end{equation}%
The elements of $\mu =\{\mu _{0},\mu _{1},...,\mu _{r}\}^{\prime }$ are
complicated functions of the elements of $\gamma =\{\gamma _{0},\gamma
_{1},...,\gamma _{r}\}^{\prime }$, $\alpha $ and $b_{1}$,...,$b_{k}$, and%
\begin{equation}
\rho =\sum_{j=1}^{k+1}\rho _{j}\text{, }\beta _{j}=-(\rho _{j+1}+...+\rho
_{k+1})\text{, }j=1,...,k\text{, }p=k+1\text{.}  \tag{9}
\end{equation}%
In fact only for a full trend polynomial, $\gamma ^{\prime }x_{t}-\rho
_{1}\gamma ^{\prime }x_{t-1}-...-\rho _{p}\gamma ^{\prime }x_{t-p}$ reduces
to $\mu ^{\prime }x_{t}$ after manipulation. This invariance does not hold
for any other set of regressors in $x_{t}$, or if some time powers are
missing from $x_{t}$. However, researchers may incorrectly assume the
invariance is correct for all variables. They may falsely employ it,
creating what we shall call "spurious efficiencies", by throwing important
variables out of the DF autoregression. To make it clear, Eq. (8) is only
valid for a trend polynomial. Eq. (7) is correctly specified, but it may be
cumbersome to make it feasible. To give an example of potential
misspecification, note that Eq. (7) dictates that lags of the break dummy
variables must be included in the DF autoregression. Perron (1989) ignores
this fact, although Kim and Perron (2009) correct this mistake. In addition,
the DF approach loses the first $p$ observations of the autoregression, and
there is no obvious way to recover the information they contain.

\section{Two Step and Extended Two Step DF Autoregression}

\noindent Hopefully, there is an alternative, although neglected, approach
to calculate DF and DF-type tests. Note that although the method is known,
it has never been employed for unit root testing.\footnote{%
Only an inefficient variant of the method is sometimes used in the
literature.} We term it the two step approach. It relies on work by Durbin
(1970), Breusch (1978), and Godfrey (1978 a, b) and (1988), see also
Davidson and MacKinnon (1993). To this end, from Eq. (1) and (5), one
obtains the infeasible regression%
\begin{equation}
y_{t}=\gamma ^{\prime }x_{t}+\sum_{j=1}^{p}\rho _{j}z_{t-j}+\varepsilon _{kt}%
\text{, }t=p+1,...,T\text{.}  \tag{10}
\end{equation}%
For feasibility, we apply LS to Eq. (1) to obtain estimator $\hat{\gamma}$
for $\gamma $, and (current and lagged) residual $\hat{z}_{t-j}=y_{t-j}-\hat{%
\gamma}^{\prime }x_{t-j}$ or $z_{t-j}=(\hat{\gamma}-\gamma )^{\prime
}x_{t-j}+\hat{z}_{t-j}$\ for $j=0,1,...,p$. Secondly, substituting these
relationships into Eq. (10), one obtains the feasible form%
\begin{eqnarray}
y_{t} &=&\gamma ^{\prime }x_{t}+\sum_{j=1}^{p}\rho _{j}\hat{z}%
_{t-j}+\varepsilon _{kt}^{\ast }=\gamma ^{\prime }x_{t}+\rho \hat{z}%
_{t-1}+\sum_{j=1}^{k}\beta _{j}\Delta \hat{z}_{t-j}+\varepsilon _{kt}^{\ast }%
\text{,}  \TCItag{11} \\
\varepsilon _{kt}^{\ast } &=&\sum_{j=1}^{p}\rho _{j}(\hat{\gamma}-\gamma
)^{\prime }x_{t-j}+\varepsilon _{kt}\text{, }t=p+1,...,T\text{.}  \notag
\end{eqnarray}%
Eq. (11) is always correctly specified, and delivers the DF test. When the
feasible version of Eq. (7) is correctly specified, Eq. (11) and (7) give
identical values for the DF test. This is so because both methods provide
alternative reduced form specifications of the same structural model of Eq.
(1) and (5). However, only Eq. (11) is always correctly specified, and this
is the reason we recommend it. An additional feature of Eq. (11) is that it
transforms to%
\begin{eqnarray}
\hat{z}_{t} &=&(\gamma -\hat{\gamma})^{\prime }x_{t}+\sum_{j=1}^{p}\rho _{j}%
\hat{z}_{t-j}+\varepsilon _{kt}^{\ast }=  \TCItag{12} \\
&&(\gamma -\hat{\gamma})^{\prime }x_{t}+\rho \hat{z}_{t-1}+\sum_{j=1}^{k}%
\beta _{j}\Delta \hat{z}_{t-j}+\varepsilon _{kt}^{\ast }\text{,}  \notag
\end{eqnarray}%
after subtracting $\hat{\gamma}^{\prime }x_{t}$ from both of its sides. Eq.
(12)/(11) is the correctly specified residual based autoregression to
calculate the DF test. Sometimes, an inefficient variant of the "DF" test is
calculated from an autoregression similar to Eq. (12) but with $x_{t}$
omitted. It is easy to show inefficiency. The employed misspecified
autoregression is%
\begin{equation}
\hat{z}_{t}=\sum_{j=1}^{p}\rho _{j}\hat{z}_{t-j}+\varepsilon _{kt}^{\ast
\prime }=\rho \hat{z}_{t-1}+\sum_{j=1}^{k}\beta _{j}\Delta \hat{z}%
_{t-j}+\varepsilon _{kt}^{\ast \prime }\text{, }\varepsilon _{kt}^{\ast
\prime }=(\gamma -\hat{\gamma})^{\prime }x_{t}+\varepsilon _{kt}^{\ast }%
\text{.}  \tag{13}
\end{equation}%
It is apparent that%
\begin{equation}
var(\varepsilon _{kt}^{\ast \prime })>var(\varepsilon _{kt}^{\ast })\text{.}
\tag{14}
\end{equation}%
This is true because $\gamma -\hat{\gamma}$ is stochastic. Hence, omitting $%
x_{t}$ from Eq. (12) increases the error variance of the resulting
regression, and this misspecification induces inefficiency in finite samples.%
\footnote{%
This is well known and avoided in the literature of serial correlation
testing (the Breusch-Godfrey test). Unfortunately, unit root testing
literature ignores this and employes inefficient estimators.} Note that this
inefficiency does not vanish asymptotically, if \thinspace $z_{t}$ has a
unit root and $x_{t}$ contains an intercept. It is known (see Durlauf and
Phillips (1988)) that, in this case, the intercept estimator in $\hat{\gamma}
$ is inconsistent. Although Eq. (11) and (12) are identical, Eq. (11) is to
be preferred. This is because it identifies and estimates the structural
parameters $\gamma $ (better than $\hat{\gamma}$), since the LS estimator
for $\gamma $, say $\tilde{\gamma}$, takes into account all serial
correlation. We may denote $\tilde{\rho}$ the corresponding estimator for $%
\rho $. Eq. (11) definitely provides the correct DF test, avoiding
misspecification. Nevertheless, it also ignores the first $p$ observations,
neglecting useful information.\newline
\newline
Either for robustification, or further efficiency gains, or even at the cost
of some minimal efficiency loss, we propose to augment Eq. (11) with the
first $p$ observations. This comes from employing zero padded, instead of
truncated, lags in estimation. Belsley (1996) provides a Monte Carlo study
that uses zero padded lags in serial correlation testing. We increase the
sample in Eq. (11) (or (12)) from $T-p$ to $T$ by zero padding the lags.
That is by using $0$ in place of the missing values $\hat{z}_{0}$,$\hat{z}%
_{-1}$,...,$\hat{z}_{-p+1}$. Zero padding (and increasing the sample size)
is expected to do very little efficiency harm. Nevertheless, it allows
useful information about the initial observations to be utilised, via
(roughly) employing the first $p$ rows of Eq. (1). We thus recommend the
following regression for calculating the new or efficient DF test:%
\begin{eqnarray}
y_{t} &=&\gamma ^{\prime }x_{t}+\sum_{j=1}^{p}\rho _{j}\hat{z}%
_{t-j}+\varepsilon _{kt}^{\ast \ast }  \TCItag{15} \\
&=&\gamma ^{\prime }x_{t}+\rho \hat{z}_{t-1}+\sum_{j=1}^{k}\beta _{j}\Delta 
\hat{z}_{t-j}+\varepsilon _{kt}^{\ast \ast }\text{, }\hat{z}_{0},\hat{z}%
_{-1},...,\hat{z}_{-p+1}=0\text{, }t=1,...,T\text{.}  \notag
\end{eqnarray}%
We give an example to clarify our discussion above. Perron and Yabu (2009),
among others, employ $y_{t}=\gamma _{0}+\gamma _{1}t+z_{t}$ with $%
z_{t}=\alpha z_{t-1}+u_{t}$ and $u_{t}$ iid. The LS estimator for $\alpha $, 
$\hat{\alpha}$, is obtained from the autoregression $\hat{z}_{t}=\alpha \hat{%
z}_{t-1}+u_{t}^{\prime }$, $t=2,...,T$. ($\hat{z}_{t}$ is the LS residual of
the DGP.) This estimator is inefficient. The efficient estimator must be
obtained from either $y_{t}=\gamma _{0}+\gamma _{1}t+\alpha \hat{z}%
_{t-1}+u_{t}^{\prime \prime }$ or equivalently $\hat{z}_{t}=\gamma
_{0}^{\prime }+\gamma _{1}^{\prime }t+\alpha \hat{z}_{t-1}+u_{t}^{\prime
\prime }$, $t=2,...,T$. (Note that $\gamma _{0}^{\prime }$ and $\gamma
_{1}^{\prime }$, in the latter regression, do not estimate true $\gamma _{0}$
and $\gamma _{1}$, only the former regression does this. In fact, $\gamma
_{0}^{\prime }=\gamma _{0}-\hat{\gamma}_{0}$ and $\gamma _{1}^{\prime
}=\gamma _{1}-\hat{\gamma}_{1}$, where $\hat{\gamma}_{0}$ and $\hat{\gamma}%
_{1}$ are DGP LS estimators for $\gamma _{0}$ and $\gamma _{1}$,
respectively.) (Also it is true that $u_{t}^{\prime }=\gamma _{0}^{\prime
}+\gamma _{1}^{\prime }t+u_{t}^{\prime \prime }$.) The resulting estimator
for $\alpha $, from both preceding regressions, denoted $\hat{\alpha}_{TR}$,
is the efficient estimator. (Both regressions give common standard error for 
$\hat{\alpha}_{TR}$.) Note that this estimator is identical to the DF
estimator $\hat{\alpha}_{DF}$ ($\hat{\alpha}_{TR}\equiv \hat{\alpha}_{DF}$)
obtained from the DF autoregression $y_{t}=\mu _{0}+\mu _{1}t+\alpha
y_{t-1}+u_{t}$, $t=2,...,T$. The same is true for its standard error. That
is the DF estimator is efficient. Since all discussed estimators are derived
from autoregressions that lose the first observation, we propose the zero
padded estimator $\hat{\alpha}_{ZP}$ for $\alpha $. Setting $\hat{z}_{0}=0$,
either $y_{t}=\gamma _{0}+\gamma _{1}t+\alpha \hat{z}_{t-1}+u_{t}^{\prime
\prime }$ or equivalently $\hat{z}_{t}=\gamma _{0}^{\prime }+\gamma
_{1}^{\prime }t+\alpha \hat{z}_{t-1}+u_{t}^{\prime \prime }$, $t=1,...,T$,
delivers $\hat{\alpha}_{ZP}$. This way the first observation is retained via 
$y_{1}=\gamma _{0}+\gamma _{1}+u_{1}^{\prime \prime }$ and robustifies
inference. (This approach cannot be applied to the DF autoregression.) Note
that $\hat{\alpha}_{ZP}$ and $\hat{\alpha}_{DF}$/$\hat{\alpha}_{TR}$ have
very close efficiency properties. However, only $\hat{\alpha}_{ZP}$ employes
the first observation, and is recommended.\newline
\newline
Eq. (11), or (12), definitely improves efficiency over Eq. (13). However,
Eq. (15) must be recommended, even if it is slightly less efficient than Eq.
(11) or (12). This is so because of its utilisation of the initial
observations. There is no need to prove theoretically that zero padding
increases efficiency. The recommendation is on grounds of utilisation of the
first $p$ observations. We denote Eq. (15) estimators for $\gamma $ and $%
\rho $ as $\bar{\gamma}$ and $\bar{\rho}$, respectively.\newline
\newline
We focus on Eq. (15), which gives the efficient DF test, but this also
applies to Eq. (11), which gives the correctly specified, original/usual DF
test. The DF $t$ test is the signed squared root of the Wald/$F$ test for
excluding $\hat{z}_{t-1}$ from the regression%
\begin{eqnarray}
y_{t}-\hat{z}_{t-1} &=&\gamma ^{\prime }x_{t}+(\rho -1)\hat{z}%
_{t-1}+\sum_{j=1}^{k}\beta _{j}\Delta \hat{z}_{t-j}+\varepsilon _{kt}^{\ast
\ast }\text{,}  \TCItag{16} \\
\hat{z}_{0},\hat{z}_{-1},...,\hat{z}_{-p+1} &=&0\text{, }t=1,...,T\text{.} 
\notag
\end{eqnarray}%
All tests: Wald/$F$, LR, and LM have the same size controlled finite sample,
size adjusted power. However, they may give different decision, when applied
to an empirical time series. It is of importance to construct ($t$ versions)
of the LR and LM versions of the unit root test. This is so because the
number of excluded variables is one. We denote the DF $t$ test of Eq.
(15/16) as the $t_{DF}^{\ast }$. Since the LR test has properties close to
the Wald test, we only focus on the LM test. The new LM $t$ test corrects
for the sample size and the number of variables included in the
autoregression. We denote the new LM $t$ test as $t_{LM}^{\ast }$. Similar
notation, without the star, is assigned to the corresponding tests of Eq.
(11). Let $F$ denote the $F$-test for $\rho -1=0$ in Eq. (15/16) (the $F$
test for the exclusion of $\hat{z}_{t-1}$). The $t_{DF}^{\ast }$ is also
derived as%
\begin{equation}
t_{DF}^{\ast }=sign(\bar{\rho}-1)F^{1/2}  \tag{17}
\end{equation}%
where $\bar{\rho}$ is the LS estimator of $\rho $. There is a clear, well
known relationship between the Wald/$F$ test, $F$, and the LM test, denoted $%
\chi $. It is well known that%
\begin{equation}
F=(T-m)\chi /(T-\chi )\text{,}  \tag{18}
\end{equation}%
where $m$ is the total number of regressors in Eq. (15/16). Solving for $%
\chi $, results in%
\begin{equation}
\chi =\frac{TF}{(T-m)+F}\text{.}  \tag{19}
\end{equation}%
The corresponding $t$ version of the new LM test, $t_{LM}^{\ast }$, is%
\begin{equation}
t_{LM}^{\ast }=sign(\bar{\rho}-1)\chi ^{1/2}=sign(\bar{\rho}-1)\{\frac{TF}{%
(T-m)+F}\}^{1/2}\text{.}  \tag{20}
\end{equation}%
Note that as $T$ gets large, $t_{DF}^{\ast }$ and $t_{LM}^{\ast }$ become
close. That is asymptotically, the two tests are going to be the same, but
not in finite samples.

\section{Conclusions}

\noindent We have revisited estimation theory for (parametric) unit root
testing. The original/usual DF test, obtained from the one step approach, is
quite likely to be subject to potential misspecification. Especially, when a
complicated trend function is to be employed. In addition, it cannot retain
the first observations. An inefficiency criticism applies to rough
calculation of the "DF" test from an autoregression of the LS residuals of
the DGP (that does not include the original levels regressors). This
criticism is shown to be true, after demonstrating and exemplifying the
correct two step procedure to obtain the DF test that correctly utilises LS
residuals. This neglected/new method is always correctly specified, it gives
the original DF test, and provides inference about the structural parameters
of the DGP. (It is an alternative for the original, one step DF approach.)
Furthermore, we also propose to improvise this method by (roughly) employing
the first $p$ rows of the DGP. This is done by zero padding, instead of
truncating, lagged residuals. The sample size increase and utilisation of
the first $p$ rows of the DGP robustify estimation and inference. No or very
minimal efficiency harm is to be expected by zero padding. Or even if there
is some minor efficiency harm, there is associated size robustness of
resulting tests to compensate. This gives what we call the efficient/new DF
or DF-type test, with associated new DF autoregression.


\begin{thebibliography}{99}
\bibitem{AG1992} Agiakloglou, C. and Newbold, P. (1992) Empirical Evidence
on Dickey-Fuller Type Tests, \textit{Journal of Time Series Analysis}, 
\textbf{13}, 471--83.

\bibitem{Belsley} Belsley, D. A. (1996) Doing Monte Carlo Studies with
Mathematica\registered , Chapter 15 in H. R. Varian (ed.), Computational
Economics and Finance, Modeling and Analysis with Mathematica\registered ,
Springer-Verlag, New York.

\bibitem{ChangPark} Chang, Y. and Park, J. Y. (2002) On the Asymptotics of
ADF Tests for Unit Roots, \textit{Econometric Reviews}, \textbf{21}, 431-447.

\bibitem{Breusch1} Breusch, T. S. (1978) Testing for Autocorrelation in
Dynamic Linear Models, \textit{Australian Economic Papers}, \textbf{17},
334-355.

\bibitem{Davidson} Davidson, R. and MacKinnon, J. G. (1993) \textit{%
Estimation and Inference in Econometrics}, Oxford University Press, Oxford.

\bibitem{DNSW1992a} DeJong, D., Nankervis, J., Savin, N., and Whiteman, C.
(1992 a) Integration versus Trend Stationarity in Macroeconomic Time Series, 
\textit{Econometrica}, \textbf{60}, 423-434.

\bibitem{DNSW1992b} DeJong, D., Nankervis, J., Savin, N., and Whiteman, C.
(1992 b) The Power Problems of Unit Root Tests for Time Series with
Autoregressive Errors, \textit{Journal of Econometrics}, \textbf{53},
323-343.

\bibitem{df1} Dickey, D. A. and Fuller, W. A. (1979) Distribution of the
Estimators for Autoregressive Time Series with a Unit Root, \textit{Journal
of the American Statistical Association}, \textbf{74}, 427-431.

\bibitem{df2} Dickey, D. A. and Fuller, W. A. (1981) Likelihood Ratio
Statistics for Autoregressive Time Series with a Unit Root, \textit{%
Econometrica}, \textbf{49}, 1057-1072.

\bibitem{Dickey1} Dickey, D. A. and Said, S. E. (1981) Testing ARIMA(p, 1,
q) against ARMA(p+1, q), \textit{Proceedings of the American Statistical
Association, Business and Economic Statistics Section}, \textbf{28}, 318-322.

\bibitem{Durbin2} Durbin, J. (1960) Estimation of Parameters in Time-Series
Regression Models, \textit{Journal of the Royal Statistical Society}, Series
B, \textbf{22}, 139-153.

\bibitem{Durbin1} Durbin, J. (1970) Testing for Serial Correlation in Least
Squares Regression when some of the Regressors are Lagged Dependent
Variables, \textit{Econometrica}, \textbf{38}, 410-421.

\bibitem{DP1988} Durlauf, S. and Phillips, P. C. B. (1988) Trends versus
Random Walks in Time Series Analysis, \textit{Econometrica}, \textbf{56},
1333-1354.

\bibitem{Fuller1996} Fuller, W. A. (1996) \textit{Introduction to
Statistical Time Series}, 2nd ed., (Wiley, New York).

\bibitem{Godfrey2} Godfrey, L. G. (1978 a) Testing Against General
Autoregressive and Moving Average Error Models when the Regressors Include
Lagged Dependent Variables, \textit{Econometrica}, \textbf{46}, 1293-1302.

\bibitem{Godfrey3} Godfrey, L. G. (1978 b) Testing for Higher Order serial
Correlation in Regression Equations when the Regressors Include Lagged
Dependent Variables, \textit{Econometrica}, \textbf{46}, 1303-1310.

\bibitem{Godfrey1} Godfrey, L. G. (1988) \textit{Misspecification Tests in
Econometrics (The Lagrange Multiplier Principle and other Approaches)},
Cambridge University Press: Cambridge.

\bibitem{Harvey2009} Harvey, D. I., Leybourne, S. J., and Taylor, A. M. R.
(2009) Unit Root Testing in Practice: Dealing with Uncertainty over the
Trend and Initial Condition, \textit{Econometric Theory}, \textbf{25},
587-636.

\bibitem{JanssonNielsen2012} Jansson, M. and Nielsen, M. O. (2012) Nearly
Efficient Likelihood Ratio Tests of the Unit Root Hypothesis, \textit{%
Econometrica}, \textbf{80}, 2321-2332.

\bibitem{KP2009} Kim, D. and Perron, P. (2009) Unit Root Tests allowing for
a Break in the Trend Function at an Unknown Time under both the Null and
Alternative Hypotheses, \textit{Journal of Econometrics}, \textbf{148}, 1-13.

\bibitem{Leubourne1999} Leybourne, S. J. and Newbold, P. (1999) The
behaviour of Dickey-Fuller and Phillips-Perron tests under the alternative
hypothesis, \textit{Econometrics Journal}, \textbf{2}, 92-106.

\bibitem{Leybourne2000} Leybourne, S. J. and Newbold, P. (2000) Behaviour of
the standard and symmetric Dickey-Fuller type tests when there is a break
under the null hypothesis, \textit{Econometrics Journal}, \textbf{3}, 1-15.

\bibitem{Nelson3} Nelson, C. R. and Plosser, C. I. (1982) Trends and Random
Walks in Macroeconomic Time Series, \textit{Journal of Monetary Economics}, 
\textbf{10}, 139-162.

\bibitem{Perron1989} Perron, P. (1989) The Great Crash, the Oil Price Shock,
and the Unit Root Hypothesis, \textit{Econometrica}, \textbf{57}, 1361-1401.

\bibitem{py2009a} Perron, P. and Yabu, T. (2009 a) Estimating Deterministic
Trends with an Integrated or Stationary Noise Component, \textit{Journal of
Econometrics}, \textbf{151}, 56-69.

\bibitem{Said1} Said, S. E. and Dickey, D. A. (1984) Testing for Unit Roots
in Autoregressive-Moving Average Models of Unknown Order, \textit{Biometrika}%
, \textbf{71}, 599-607.

\bibitem{Said2} Said, S. E. and Dickey, D. A. (1985) Hypothesis Testing in
ARIMA(p,1,q) Models, \textit{Journal of the American Statistical Association}%
, \textbf{80}, 369-374.

\bibitem{schwert2} Schwert, G. W. (1989) Tests for Unit Roots: A Monte-Carlo
Investigation, \textit{Journal of Business and Economic Statistics}, \textbf{%
7}, 147-159.

\bibitem{Stock1991} Stock, J. H. (1991) Confidence Intervals for the Largest
Autoregressive Root in US Macroeconomic Time Series, \textit{Journal of
Monetary Economics}, \textbf{28}, 435-459.
\end{thebibliography}
\end{document}